\documentstyle[12pt]{article}
\topmargin  - 1cm
\begin{document}
\thispagestyle{empty}
\phantom{.}
\vskip 2cm
\begin{center}
 {\bf { \Large Geometry of Manifolds and Dark Matter  \\  }
\vspace*{1cm}
{  Ivanhoe Pestov   \\  }
\vspace*{.5cm}
{\sl  Joint Institute
for  Nuclear Research,\\ Dubna, Moscow Region,141980 RUSSIA\\
e-mail: pestov@thsun1.jinr.ru \\}
\vspace*{1cm}
}
\end{center}
\begin{abstract}
It is shown that the theory of dark matter can be derived from the first
principles. Particles
representing a new form of matter  gravitate but do not interact
electromagnetically, strongly and weakly with the known elementary particles.
Physics of  these particles is defined by the Planck scales.
  \end{abstract}
\vskip  2cm
%\newpage
\section {Introduction}

The problem of invisible mass [1],[2] is acknowledged  to be among the
greatest puzzles of modern cosmology and theory of elementary particles.
The most direct evidence for the existence of large quantities of dark
matter in the Universe comes from the astronomical observation of the motion
of visible matter in galaxies [3]. One does neither know the identity
of the dark
matter nor whether there is one or more type of its structure elements.
The most commonly discussed theoretical elementary particle candidates are
a massive neutrino, a sypersymmetric neutralino and the axion. It is
self-evident that existence of such particles is a great question.

So, at present time there is a good probability that the set of elementary
particles is by no means limited by those particles, the evidence for
whose is proved by experience. Moreover, we are evidently free
to look for more deep reasons for the existence of new particles unusual
in many respects. Here the theory of structure elements of the so-called
dark matter is derived from the first principles and all properties of these
particles are listed at the end.

 \section { Geometrical framework }

According to the modern standpoint a fundamental physical theory is
the one that possesses a mathematical representation whose elements
are smooth manifold and geometrical objects defined on this manifold.
Most physicists today consider a theory be fundamental only if it
does make explicit use of this concept. It is thought that curvature
of the manifold itself provides an explanation of gravity. Within
the manifold, further structure are defined-vector fields,
connexions, particle path and so forth- and these are taken to
account for the behavior of physical world. This picture is so
generally accepted and it is based on such a long history of physical
research, that there is no reason to question it.

Geometry on
manifold M with a local coordinate system $x^i, \quad i=0,1,2,3,$ is
defined by the metric $g_{ij}$ and linear (affine) connexion
${\Gamma}^j_{ik}.$ Tensor field $g_{ij}$ is symmetric, $g_{ij} =
g_{ji},$  but linear connexion ${\Gamma}^j_{ik} $ is nonsymmetric
with respect to the covariant indices, $ {\Gamma}^j_{ik} \neq
{\Gamma}^j_{ki} $ in the general case and in any way does not link
with the metric $g.$ Really, these notions define, on a
manifold M, different geometric operations. Namely, a metric on a
manifold defines at every point the scalar product of vectors from
the tangent space and linear connection gives the translation along
any path on M.

  \section  {Symmetry}

Symmetry on a manifold can be introduced as follows.
Consider the Einstein law of gravity $R_{ij}= 0.$  Let $g_{ij}$ be a solution
of the equations $R_{ij}= 0 $  and $f^{i}(x)$ are four functions defined by
the demand that all expressions in the formula
$$\tilde{g}_{ij}(x)= f^k_i(x) f^l_j(x) g_{kl}(f(x)), $$ where $
f^k_{i}(x)= \partial_{i}f^{k}(x),$ have the meaning. It can be shown
that $\tilde{g}_{ij}(x)$ is a symmetric tensor and a new solution of
the Einstein eqs. $R_{ij}= 0.$ The point transformation $x^i
\Rightarrow f^{i}(x)$  is  a local diffeomorphism.  All such
transformations form a group of local diffeomorphisms. As it follows
from the above statement, the group thus defined is a symmetry group of
gravitation interactions. It should be noted that it is the most general
group of coordinate transformations.

Another very general group of transformations on M can be defined as
follows. Consider the most general linear transformation of vector fields
 $$ {\bar V}^i = S^i_j V^j ,$$
 where ${\bar V}^i$  and $V^i$  are
components of vector fields, $S^i_j$ are  components of a tensor
field of the type $(1,1),$ \quad ${\rm det} (S^i_j) \neq 0.$
Of two tensor fields $S^i_j$ and $T^i_j$ of the type (1,1)
a tensor field $P^i_j = S^i_{k} \,T^k_j $ of the type (1,1) may be
constructed, called their product. With the operation of multiplication
thus defined, the set of tensor fields of the type (1,1) with
a nonzero determinant
forms a group, denoted by $G_i.$ This is a most general  gauge group on
a manifold. It can be shown that the diffeomorphism group
is the group of external automorphisms of the gauge group, i.e.
the gauge group is invariant under the transformations of the
group ${\rm Diff}(M).$ Thus, we have a nontrivial unification of
these symmetries.

The tensor fields can transform under the gauge transformations in
different ways. We will say that a tensor field $T$ of the type $(m,n)$ has
the gauge type $(p,q)$ if under the transformations of the gauge group there
is the correspondence $$ \bar T = {\underbrace{ S  \ldots S }_{p}} T
{\underbrace{ S^{-1}  \ldots  S^{-1}}_{q} },$$ where $ 0 \leq p \leq m,$ \quad
$0 \leq q \leq n,$  and $S^{-1}$ is the transformation inverse to $S,\quad
S^{-1}= (T^i_j),\quad S^i_k T^k_j = {\delta}^i_j.$ From the  equations
$R_{ij} = 0$ which  express the Einstein law of gravity  one can find
 that the Einstein gravitational potentials $g_{ij}$ have the gauge type
 $(0,0).$

  \section { Main Conjecture}

Now, we have the possibility to put forward the idea that the group of gauge
symmetry $G_i$ defines  properties of a new form of matter called the
dark matter.
From the theory of linear vector spaces and for the reasons
of symmetry and simplicity it follows that there is a {\bf single quantity}
which
can be put in correspondence with  hypothetical  particles of this matter.
This quantity is a tensor field of the type $(1,1) ,\quad
{\Psi}^i_j,$  and the gauge type $(1,1).$ Since under the action of the gauge
group a tensor field $\Psi$ is transformed as follows $$ \bar {\Psi} =
S{\Psi}S^{-1},$$ then scalars ${\Psi}^i_i = Tr \Psi$  and ${\Psi}^i_j
{\Psi}^j_i = Tr ({\Psi}{\Psi})$ are evidently invariants of the gauge
group $G_i$. It is known from the theory  of linear operators that there also
exist other invariants, but in what follows we will only use the invariant
$Tr({\Psi}{\Psi}).$

To derive the nontrivial gauge invariant equations for  $\Psi,$
first of all investigate the link between the linear connection
${\Gamma}^j_{ik}.$ and gauge group $G_i.$ To this end consider the properties
of covariant derivative with respect to linear connection ${\Gamma}^j_{ik}$
from the standpoint of  gauge symmetry in question.
The covariant derivative
of $\Psi$ with respect to the affine connection ${\Gamma}_i =
({\Gamma}^j_{ik})$ can be written in the form $$ \nabla_i{\Psi} =
\partial_i {\Psi} + [{\Gamma}_i, \Psi].$$ The use of the matrix notation is
rather evident and does not require special explanations.  Let $\bar
{\Gamma}_i = (\bar {\Gamma}^j_{ik})$  is another affine connection on $M$
and $\bar{\nabla}_i$  denotes the covariant derivative with respect to this
connection. Then $$ \bar{\nabla}_i {\Psi} = \nabla_i {\Psi} +
 [\delta{\Gamma}_i, \Psi],$$ where $\delta {\Gamma}_i = \bar {\Gamma}_i -
\Gamma_i.$  Substituting, into the above relation,  $\bar {\Psi} =
S{\Psi}S^{-1} $ instead of $\Psi,$  we get $$ \bar{\nabla_i} {\bar {\Psi}} = S
(\nabla_i {\Psi}) S^{-1} + [\delta{\Gamma}_i - S{\nabla}_i S^{-1}, \Psi ].$$
From this it follows that
\begin{equation}
 \bar {\nabla}_i \bar {\Psi} = S (\nabla_i {\Psi}) S^{-1}
\end{equation}
if $\delta {\Gamma}_i = S \nabla _i S^{-1},$ or that is the same,
\begin{equation}
	\bar {\Gamma}_i = \Gamma_i + S \nabla_i S^{-1}.
\end{equation}

Let
$$ ({B_{ij}}^k_l) = B_{ij} = \partial_i {\Gamma}_j - \partial_j {\Gamma}_i + [\Gamma_i, \Gamma_j]$$
be the Riemann tensor of the affine connection $\Gamma_i$, then from (2) it
follows that \begin{equation} \bar B_{ij}  = S B_{ij} S^{-1}, \end{equation}
where $\bar{B}_{ij}$ is the Riemann tensor of the connection $\bar {\Gamma}_i.$
Thus, from (2) and (3) it follows that in the framework of the gauge group
$G_i$ one can consider the affine connection ${\Gamma}^j_{ik}.$ as the
gauge field
and the tensor $B_{ij}$ as the strength tensor of this field. According to (1)
a tensor field $\nabla_i {\Psi}$ has the same gauge type as $\Psi,$  but this is
not true for the second covariant derivative of ${\Psi}$ or $B_{ij}.$ For this
reason it is necessary to introduce the important notion of the gauge
covariant derivative, which  does not
change the gauge type of the quantity in question.

Let $T$ be a tensor field (tensor density) of the gauge type $(1,1),$  then
by definition
$$ D_i T = \partial_i T + [\Gamma_i , T]$$
is the gauge covariant derivative. For example, for the Riemann tensor we have
$$ D_i B_{jk} = \partial_i B_{jk} + [\Gamma_i , B_{jk}].$$
As it must be, for the field $\Psi$   the gauge covariant derivative
coincides with the standard covariant derivative, $ D_i {\Psi} = \nabla_i
{\Psi}.$ In the general case the operator $D_i$ is not general covariant,
since $D_i T$ will not  always be a tensor field together with $T.$
However, the commutator $[D_i , D_j]$ is always general covariant, because $$
[D_i , D_j] T = [B_{ij} ,T].$$ From this we get the important relation for the
Riemann tensor \begin{equation} [D_i , D_j ] B_{kl} = [B_{ij} , B_{kl}]
\end{equation}

Thus, the gauge symmetry shows  that not only the Einstein gravitational
potentials $g_{ij}$ but also the tensor field $\Psi$ and gauge field
$\Gamma_i$ are to be treated as primary fields tightly connected with the
symmetry and geometry on the  manifold. Before writing the simplest
gauge
invariant equations, that express the law of interaction of these fields,
it must be noted that all the three fields have  geometrical
interpretation.  For the field $g_{ij }$ it is well known, so we dwell
only on the geometrical interpretation of the fields $\Psi$  and
$\Gamma,$ considered as a whole.

Let $\delta V^i = dV^i + \Gamma ^i_{jk} {\dot x}^j V^k  dt  $  be an
infinitesimal change of the vector field $V$ on a curve $\gamma(t)$  and $
S^i_j = \delta ^i_j + \Psi ^i_j dt$  be an infinitesimal gauge transformation.
We assume that at every moment of time $t$ an infinitesimal change of a vector
$V$  along a curve $\gamma(t)$  is equal to an infinitesimal linear
transformation of the vector field $V$ induced  by the gauge group, $ \delta
V^i = \Psi ^i_j V^i dt .$  Thus, we obtain a system of ordinary linear
homogeneous differential equations for  $V^i(t)$ $$ \frac{dV^i}{dt} +
\Gamma ^i_{jk} \frac{dx^j}{dt} V^k = \Psi^i_j V^j.$$ from which it follows
that a composite geometrical object $ (\Psi , \Gamma_i)$ defines the general
law of translations of a vector field along the given curve
$\gamma(t).$

 \section   { Field Equations }

Now we shall assume that the field $\Psi$ has the gauge type (1,1),
under the action of the gauge group the $\Gamma$- field is transformed
by the law (2) and construct a $G_i$-invariant theory of the interaction of
those fields.
The simplest gauge invariant Lagrangian of the field $\Psi$  has the  form
\begin{equation}
L_{\Psi}  = - \frac{1}{2} Tr (D_i {\Psi} D^i {\Psi} - m^2 {\Psi}{\Psi} ),
\end{equation}
where $m$ is a constant, $D_i = g^{ij} D_j .$   From (5)  by the variation with
respect to $\Psi$  we obtain the following equations
\begin{equation}
	 D_i ( \sqrt{|g|} D^i {\Psi}) + m^2 \sqrt{|g|} {\Psi} = 0 ,
\end{equation}
where $ |g| $ is the absolute value of the determinant of the matrix $ (g_{ij}).$
When deriving (6) one should  take into account that $Tr (D_i {\Psi}) = \partial_i(Tr{\Psi}).$
In accordance with (6) one can consider $m$  as the mass of a particle
defined by the field $\Psi.$  The simplest Lagrangian of the gauge field $\Gamma$
is a direct consequence of  (3)
\begin{equation}
	L_{{\Gamma}} = - \frac{1}{4} Tr (B_{ij}  B^{ij}),
\end{equation}
where  $ B^{ij}= g^{ik}g^{jl} B_{kl}.$  Varying the Lagrangian $ L =
L_{{\Psi}} + L_{{\Gamma}}$ with respect to $\Gamma$  with the help of the
relation $\delta B_{ij} = D_i {\delta} {\Gamma}_j - D_j {\delta}{\Gamma}_i$ we
obtain the following equations of the gauge field $\Gamma$ \begin{equation}
  D_i (\sqrt {|g|} B^{ij}) = \sqrt{|g|} J^j,
\end{equation}
the right hand side of which contains the tensor field of the third rank
\begin{equation}
	    J^i = [\Psi , D^i {\Psi}].
\end{equation}
This field obviously has the gauge type $(1,1).$  The tensor current $J^i$
has to satisfy the equation
\begin{equation}
       D_i (\sqrt{|g|} J^i) = 0,
\end{equation}
as in accordance with (4) , $D_i D_j (\sqrt{|g|} B^{ij})  \equiv 0.$  From (6)
and (9) it follows that $J^i$ really satisfies  equation (10) and thus the
system of  equations (6) , (8) is consistent.

Varying the Lagrangian $ L = L_{{\Psi}} + L_{{\Gamma}}$  with respect
  to $g^{ij}$ we obtain the so-called metric tensor of energy-momentum of the
considered system of interacting fields \begin{equation} T_{ij} = Tr (D_i
   {\Psi} D_j {\Psi}) + Tr (B_{ik} B_j{}^k) + g_{ij} L, \end{equation} where $
B_j{}^k = B_{jl} g^{kl}.$ If the fields $\Psi$ and $\Gamma$  satisfy
equations (6) and (8),  then one can show that the metric tensor of the
energy-momentum satisfies the well-known equations
$$ T^{ij}{}_{;i} = 0$$
where the semicolon denotes as usual the covariant derivative  with respect to
the Levi-Civita connection belonging to the field $g_{ij}$ $$ \{^i_{jk}\} =
\frac{1}{2}g^{il} (\partial_j g_{kl} + \partial_k g_{jl} - \partial_l
g_{jk}).$$ It is evident that the metric tensor energy-momentum is  gauge
invariant.

Now we can write down the full action for the fields
$g_{ij}, \quad {\Psi},\quad {\Gamma} $
\begin{eqnarray} S = -
\frac{c^3}{G}\int R \sqrt{|g|} d^4 x &-&\frac{\hbar}{2}\int Tr(D_i {\Psi} D^i
{\Psi} + m^2 {\Psi}{\Psi}) \sqrt{|g|} d^4 x - \nonumber\\
&-&\frac{\hbar}{4}\int Tr(B_{ij}
B^{ij}) \sqrt{|g|}d^4 x,
\end{eqnarray}
where $R$ is the scalar curvature, $G$
is the Newton  gravitational constant and $\hbar$ is the Planck constant.
From the geometrical interpretation of the fields $\Psi$ and $\Gamma$ it
follows that they have the dimension $sm^{-1}.$
As all coordinates can be considered to have the dimension $sm,$  the action
$S$ has a correct dimension. It is necessary to substantiate only why we
have introduced the Planck constant $\hbar$ into the full action $S$
and not, say, the constant of interaction $\varepsilon$
with the gauge field $\Gamma,$  similar to the electric charge of the electron
$e.$

Consider the infinitesimal transformations of the diffeomorphism group ,
$\bar x^i = x^i + K^i(x)dt.$ If under such transformations the gravitational
potentials  $g_{ij}$  do not vary, i.e. the vector field $K^i(x)$ satisfies
the Killing equations $$ K^i \partial_i g_{jl} + g_{il} \partial_j K^i +
g_{ji} \partial_l K^i = K_{j;l} + K_{l;j} = 0,$$ then the vector field $P^i =
T^{ij}K_j$ will satisfy the equation $P^i{}_{;i} = 0.$ Integrating this
equation we obtain the conservation law. The Killing equations  impose severe
constraints on the gravitational potentials. Thus, the Killing
equations are completely integrable if the tensor of curvature of the metric
$g_{ij}$ satisfies the equations:  $$ R_{ijkl} = \frac{R}{12} (g_{ik}
g_{jl} - g_{il} g_{jk}).$$ In the general case the Killing equations have no
solutions at all and there are no conservation laws. This result
allows us to understand the absence of the constant of interaction of matter
 fields  with the gravitational field similar to the electric charge $e$  and
why the  Newton  gravitational constant $G$ does not enter into the
equations of  matter fields.  It is impossible to switch on or switch
off the gravitational field.  It does not admit the existence of the
gravitational screen, the "gravitational charge" does not exist as an
invariant fundamental notion.

Now consider infinitesimal gauge transformations $ S^i_j = \delta^i_j + \Omega^i_j dt.$
If under such transformations the gauge field $\Gamma_i$  does not vary,
i.  e. the  tensor field $\Omega^i_j$ satisfies the equations $$
\partial^i{\Omega^j_k} + \Gamma^j_{il} {\Omega}^l_k - \Omega^j_l
{\Gamma}^l_{ik} = \nabla_i{\Omega}^j_k = 0,$$ then the vector field $Q^i = Tr
(J^i {\Omega}),$  where $J^i$  is a tensor current (9), will satisfy the
equation $ Q^i_{}{;i} = 0.$  Integrating this equation we obtain the
conservation law as usual. The equation $\nabla_i {\Omega}^j_k = 0,$
like the Killing equations, imposes severe constraints, in the given
case, on the gauge field $\Gamma_i.$ Thus, the equation $\nabla_i
{\Omega}^j_k = 0$ is completely integrable, if the strength tensor $B_{ij}$
satisfies the equations ${B_{ij}}^k_l = 1/4 {B_{ij}}^m_{m} {\delta}^k_l.$  So,
in the considered case, generally speaking, there are no  conservation
laws too. From this fact it is natural to conclude, that there is
no special coupling constant ( like the electric charge) of the gauge field
 $\Gamma$ with the field of matter $\Psi.$  The gauge field $\Gamma$  is
impossible to screen. In the known sense, the Planck constant $\hbar$ is
analogous to the Newton  gravitational constant $G.$
The Planck constant $\hbar$  characterizes the intensity of interactions
of the field $\Psi$ with the gauge field $\Gamma.$ Interactions in
question has a pure quantum nature. From this consideration one can
conclude that non-abelian gauge fields and gravity are not in the
scope of perturbation theory and we need new  approaches here.

 \section  { General Physical Interpretation }

Varying the full action (12) with respect to $g^{ij}$ we derive the
Einstein equations
$$ R_{ij} - \frac{1}{2} g_{ij} R = l^2 T_{ij},$$
where $l = \sqrt{\hbar G/c^3}$  is the Planck length and $T_{ij}$  is the
metric tensor of energy-momentum (11). Thus, it is shown that the problem
formulated above has the solution.
A tensor field of the second rank $\Psi^i_j$ describes unknown gravitating
particles which, as it was shown earlier, are a single  source of the gauge
field $\Gamma$ that is known as the affine connection in geometry. As it is
known, the affine connection has always played a fundamental role in all
attempts to develop General Relativity from the very start of its creation
[4],[5],[6]. The conclusion drawn here that the affine connection has a
conserved energy-momentum tensor and therefore may be a source of
gravitational field radically changes the view on that object.
In this connection  it should be emphasized
that equations (6) and (8) can be considered on
the background of the Minkowski spacetime. It is enough to suppose that
$g_{ij}$ is the Minkowski metric.

It is not difficult to show that for the field $\Psi$  there is no nontrivial
gauge invariant equations of the first order. Let $S$ be an element of the gauge
group; obviously, $(-S)$ also belongs to that group. Further, under the gauge
transformation  $\bar {\Psi} = S{\Psi} S^{-1}$ the same  transformation of the
field $\Psi$  corresponds to different elements $S$  and $(-S)$ of the gauge
group. We have remarked these properties   because they characterize the bose
particles.

The equations of motion for a test particle in external gravitational
and gauge fields have the form
$$ \frac{d^2 x^i}{ds^2} + \{ ^i_{jk}\} \frac{dx^j}{ds} \frac{dx^k}{ds} + \frac{\hbar}{mc} {\omega}^i{}_j  \frac{dx^j}{ds} = 0,$$
where $\omega_{ij} = Tr(B{ij})  = \partial_i Q_j - \partial_j Q_i$  and $Q_i = Tr(\Gamma_i) = \Gamma^k_{ik}.$
An additional gauge force acting on a test particle has the quantum nature and
the same form that the Lorentz force. However, the new force has one
essential feature which can be seen from a rather general consideration.
Taking the trace of equations (8) and taking into account  that in
accordance with (9), $Tr(J^i) \equiv 0,$  one obtains that if  $\Gamma$
satisfies equations (6) and (8), then $\omega_{ij}$  will satisfy the
equations $ \omega^{ij}{}_{;i} = 0.$ These equations are the same as the
Maxwell equations without sources. From this one can conclude that the
new gauge invariant force connected with the gauge field $\Gamma $
can neither be central nor Coulomb.

\section { Summary}

In accordance with the theory developed above, the main properties of
particles of dark matter are as follows:

1.  These particles practically do not interact with the known elementary
particles. Really, it is impossible to construct the Lagrangian of interaction
of the field $\Psi$ with other known fields being invariant with respect
to the transformations of the gauge group in question.

2. However, these particles, as it is established, interact gravitationally.

3. The spin of particles equals 1. To see this, one can remind that
any real matrix has the canonical (Jordan) form [7]. In one case, for
example, $\Psi = diag (\lambda_1, \lambda_2,\lambda_3,\lambda_4),$ where
$\lambda_i $ are  roots of the characteristic equation
$|{\Psi}-{\lambda}E| = 0.$ Besides, it should be noted that the expansion
$\Psi = \Psi - 1/4 Tr(\Psi)\, E + 1/4 Tr (\Psi)\, E = \Psi_0 +1/4 Tr (\Psi)\, E $
is gauge
invariant. Thus, the field $\Psi_0$ has three degrees of freedom
and we have deal with the Bose particles.

4. The mass of the particles is a single parameter of the theory. However,
under the natural condition $m = m_p =\sqrt{\hbar c/G} $ we shall obtain
the theory free from
parameters completely.

5. We have not deep reasons to introduce complex value wave function and hence
electric charge of particles equal zero.

So, actually all is known about unknown particles. However, experience
remains a sole criterion of the validity of a mathematical
construction for the physics. Last but not least. If the suggested physical
interpretation of the theory is true, then the evidences of the existence of
the dark matter, obtained from the astronomical observations, show that the
macroscopic quantum effects can appear not only on the scale of solid
(superconductivity) but also on the scale of the galaxies.
   
\end{document}